\documentclass[11pt,twoside]{article}
\usepackage[pdftex]{graphicx}
\usepackage{amsmath}
\usepackage{amssymb}
\usepackage{cite}

\usepackage{multicol}
\usepackage{epstopdf}

\begin{document}
\newcommand{\pst}{\hspace*{1.5em}}

\newcommand{\rigmark}{\em Journal of Russian Laser Research}
\newcommand{\lemark}{\em Volume 30, Number 5, 2009}

%\lhead[\fancyplain{\rigmark, {\em \lemark}}{\rigmark}]{\fancyplain{\rigmark, {\em \lemark}}{\lemark}}
%\chead{}\rhead[\fancyplain{}{\lemark}]{\fancyplain{}{\rigmark}}
%\plainfootrulewidth 0.4pt
\newcommand{\be}{\begin{equation}}
\newcommand{\ee}{\end{equation}}
\newcommand{\bm}{\boldmath}
\newcommand{\ds}{\displaystyle}
\newcommand{\bea}{\begin{eqnarray}}
\newcommand{\eea}{\end{eqnarray}}
\newcommand{\ba}{\begin{array}}
\newcommand{\ea}{\end{array}}
\newcommand{\arcsinh}{\mathop{\rm arcsinh}\nolimits}
\newcommand{\arctanh}{\mathop{\rm arctanh}\nolimits}
\newcommand{\bc}{\begin{center}}
\newcommand{\ec}{\end{center}}

\renewcommand{\labelenumi}{(\alph{enumi})} % Use letters for enumerate
\let\vaccent=\v % rename builtin command \v{} to \vaccent{}
\renewcommand{\v}[1]{\ensuremath{\mathbf{#1}}} % for vectors
\newcommand{\gv}[1]{\ensuremath{\mbox{\boldmath$ #1 $}}} 
% for vectors of Greek letters
\newcommand{\uv}[1]{\ensuremath{\mathbf{\hat{#1}}}} % for unit vector
\newcommand{\abs}[1]{\left| #1 \right|} % for absolute value
\newcommand{\avg}[1]{\left< #1 \right>} % for average
\let\underdot=\d % rename builtin command \d{} to \underdot{}
\renewcommand{\d}[2]{\frac{d #1}{d #2}} % for derivatives
\newcommand{\dd}[2]{\frac{d^2 #1}{d #2^2}} % for double derivatives
\newcommand{\pd}[2]{\frac{\partial #1}{\partial #2}} 
% for partial derivatives
\newcommand{\pdd}[2]{\frac{\partial^2 #1}{\partial #2^2}} 
% for double partial derivatives
\newcommand{\pdc}[3]{\left( \frac{\partial #1}{\partial #2}
	\right)_{#3}} % for thermodynamic partial derivatives
\newcommand{\ket}[1]{\left| #1 \right>} % for Dirac bras
\newcommand{\bra}[1]{\left< #1 \right|} % for Dirac kets
\newcommand{\braket}[2]{\left< #1 \vphantom{#2} \right|
	\left. #2 \vphantom{#1} \right>} % for Dirac brackets
\newcommand{\matrixel}[3]{\left< #1 \vphantom{#2#3} \right|
	#2 \left| #3 \vphantom{#1#2} \right>} % for Dirac matrix elements
\newcommand{\grad}[1]{\gv{\nabla} #1} % for gradient
\let\divsymb=\div % rename builtin command \div to \divsymb
\renewcommand{\div}[1]{\gv{\nabla} \cdot #1} % for divergence
\newcommand{\curl}[1]{\gv{\nabla} \times #1} % for curl
\let\baraccent=\= % rename builtin command \= to \baraccent
\renewcommand{\=}[1]{\stackrel{#1}{=}} % for putting numbers above =

\thispagestyle{plain}

\label{sh}

%\lfoot[\fancyplain{\ \\[1mm] \thepage}{\ \\[1mm]\thepage}]{\fancyplain{}{}}

\begin{center} {\Large \bf
		\begin{tabular}{c}
			WEIGHTED INFORMATION\\ AND WEIGHTED ENTROPIC INEQUALITIES
			\\[-1mm]
			FOR QUTRIT AND QUQUART STATES
		\end{tabular}
	} \end{center}

	\begin{center} {\bf
			Vladimir I.~Man'ko$^{1,2*}$ and Zhanat Seilov$^2$
		}\end{center}
		
		\medskip
		
		\begin{center}
			{\it
				$^1$Lebedev Physical Institute, Russian Academy of Sciences\\
				Leninskii Prospect 53, Moscow, Russia 119991
				
				\smallskip
				
				$^2$Moscow Institute of Physics and Technology (State University)\\
				Institutskii per. 9, Dolgoprudnyi, Moscow Region Russia 141700
			}
			\smallskip
			
			$^*$Corresponding author e-mail:~~~manko@lebedev.ru\\
		\end{center}

\begin{abstract}\noindent
	The notion of weighted quantum entropy is reviewed and considered for bipartite and noncomposite quantum systems. The known for the weighted entropy information inequality (subadditivity condition) is extended to the case of indivisible qudit system on example of qutrit.
	This new inequality for qutrit density matrix is discussed for different cases of weights and states.
	The role of weighted entropy is discussed in connection with studies of nonlinear quantum channels.

\end{abstract}

\medskip


\section{Introduction}

The classical probability distributions are characterized by such functionals as Shannon entropy \cite{shannon}. The entropy value corresponds to degree order in the classical system. The larger is the Shannon entropy, the larger degree of disorder in the system is. More detailed description of the probability distributions is associated with Renyi entropy \cite{renyi} and Tsallis entropy \cite{tsallis}. These entropies depend on extra parameter. The dependence on this parameter provides possibility to characterize the properties of the distribution in more detail. The generalization for entropy of classical system called weighted entropy was introduced by M.Belis and S.Guiasu in \cite{1968} and then developed in \cite{1971}. The reason to introduce this generalization was to show the qualitative aspect of measurement, whereas probabilities in Shannon entropy reflecting quantitative aspect. Weights can be distributed according to the relevance or utility of corresponding events or states. The weighted entropy depends on many extra parameters determining the chosen probability distribution.  Von Neumann entropy \cite{neumann} is one of the central concepts in quantum information theory. It shows the unpredictability of information content in quantum system. 

 The Shannon entropy of multipartite classical system satisfies known information-entropic inequalities like subadditivity condition for bipartite systems \cite{robinson1} and strong subadditivity condition for tripartite classical systems \cite{robinson2}. These conditions are also valid for quantum von Neumann entropy \cite{araki, carlen}. The Tsallis entropy of bipartite system also satisfies the subadditivity condition.  Y.Suhov and S.Zohren in \cite{suhov} introduced quantum weighted entropy as generalized concept of well known von Neumann entropy. In this generalization weight is represented by weight matrix. Authors of \cite{suhov} derived and proved main properties of new entropy: subadditivity, concavity and strong subadditivity. Big interest for us represents the property of subadditivity of quantum weighted entropy, which shows that mutual information can't be of negative values, and it was shown recently in \cite{manko2, manko3, chernega, markovich, hidden}. It was clarified that entropic-information inequalities known for composite systems are valid for systems without subsystems both in classical and quantum domains. 

The aim of our work is to review the properties  of weighted quantum entropy discussed in \cite{suhov, suhov2, suhov3} and to show that entropic inequalities for weighted quantum entropy like subadditivity condition found for tripartite systems \cite{suhov} are also valid for systems without subsystems. We demonstrated that this inequality fits particular qutrit states. 
There exists the nonlinear maps of density matrices called nonlinear quantum channels (see, e.g. \cite{europhys, puzko, manko4}). We will discuss the changes of the weighted entropic inequalities due to action of nonlinear quantum channels.  The paper is organized as follows: in Sec. 2 we review the notion of quantum weighted entropy. In Sec. 3 we consider the subadditivity property for qutrit state with diagonal density matrix. In Sec. 4 the conclusion and prospectives are presented.

\section{Quantum Weighted  Entropy}

Let $\rho$ be a density matrix and $\phi$ be a positive definite, Hermitian matrix both on Hilbert space $\mathcal H$. Then quantum weighted entropy is determined as follows

\be \label{qwe} S_\phi(\rho)\equiv-tr(\phi \rho \log\rho) ,
\ee where $\phi$ is called weight.
\\

Subadditivity property is represented by inequality 

\be \label{sub}	S_{\phi_{AB}}(\rho_{AB})\leq S_{\phi_A}(\rho_A)+S_{\phi_B}(\rho_B)
,\ee which is correct under the condition 

\be \label{cond} tr_{AB}(\phi_{AB} \rho_{AB})\geq tr_A(\phi_A \rho_A)tr_B(\phi_B \rho_B)
\ee
and in case $\rho_{AB}= \rho_A \otimes \rho_B$ it simplifies to equality. Reduced weights are defined as follows
\be \label{rm} \psi_A \rho_A = tr_B(\phi_{AB}\rho_{AB}) \ee 
and the same way for case B.

The idea of this work is to extend the weighted entropic inequalities (\ref{sub}) and (\ref{cond}) formulated above for composite systems to the noncomposite (indivisible) systems. Furthermore, we suggest to use the inequalities for studying the transformation of density matrix of qudit states due to action of nonlinear quantum channels.
\section{Subadditivity property for qutrit with diagonal density matrix}

Let us consider particular case of this property for qutrit. There are 3 states for qutrit which are supplemented with one new state with 0-probability. We introduce following designations:

\be \phi_A=	\begin{pmatrix} \phi_1 & 0 \\ 0 & \phi_2 \end{pmatrix},~~~~~~~~~  
\phi_B=	\begin{pmatrix} \chi_1 & 0 \\ 0 & \chi_2 \end{pmatrix},~~~~~~~~~
\phi_{AB}=\phi_A \otimes \phi_B=
\begin{pmatrix}
	\phi_1 \chi_1 & 0 & 0 & 0\\
	0 & \phi_1 \chi_2 & 0 & 0\\
	0 & 0 & \phi_2 \chi_1 & 0\\
	0 & 0 & 0 & \phi_2 \chi_2
\end{pmatrix};
\ee

\be \label{matrix} \rho_{AB}=\begin{pmatrix}
p_1 & 0 & 0 & 0\\
0 & p_2 & 0 & 0\\
0 & 0 & p_3 & 0\\
0 & 0 & 0 & 0
\end{pmatrix},~~~~~~~
 \rho_A=\begin{pmatrix}
p_1+p_2 & 0\\
0 & p_3
\end{pmatrix},~~~~~~~
\rho_B=\begin{pmatrix}
p_1+p_3 & 0\\
0 & p_2
\end{pmatrix}.
\ee

We define the condition (\ref{cond}) of ineqauality for qutrit:

\be \label{e1} tr_{AB}(\phi_{AB} \rho_{AB})=tr_{AB}\begin{pmatrix}
	\phi_1 \chi_1 p_1 & 0 & 0 & 0\\
	0 & \phi_1 \chi_2 p_2 & 0 & 0\\
	0 & 0 & \phi_2 \chi_1 p_3 & 0\\
	0 & 0 & 0 & 0
	\end{pmatrix}.
\ee
	
	$$tr_A(\phi_A \rho_A)tr_B(\phi_B \rho_B)
	=tr(\begin{pmatrix}
	\phi_1 & 0 \\
	0 & \phi_2 \end{pmatrix}
	\begin{pmatrix}
	p_1+p_2 & 0\\
	0 & p_3
	\end{pmatrix}) tr(\begin{pmatrix} \chi_1 & 0 \\ 0 & \chi_2 \end{pmatrix}\begin{pmatrix}
	p_1+p_3 & 0\\
	0 & p_2
	\end{pmatrix})=$$
	$$=({\it substitution~} p_3=1-p_1-p_2)=$$
\be \label{e2} =\phi_1 \chi_1(p_1+p_2-p_1p_2-p_2^2)+\phi_1 \chi_2(p_1p_2+p_2^2)+\phi_2 \chi_1(1-p_1-p_2-p_2+p_1p_2+p_2^2)+\phi_2\chi_2(p_2-p_1p_2-p_2^2).
\ee

We can write expression (\ref{cond}) in form $tr_{AB}(\phi_{AB} \rho_{AB}) - tr_A(\phi_A \rho_A)tr_B(\phi_B \rho_B)\geq 0$. After substitution from (\ref{e1}) and (\ref{e2}) we obtain new inequality
$$(p_2^2+p_2 p_1 -p_2)(\phi_1\chi_1-\phi_1\chi_2 -\phi_2 \chi_1 +\phi_2 \chi_2)\geq0~,$$	\\
$$\underbrace{p_2(1-p_2-p_1)}_{>0} (\phi_1-\phi_2)(\chi_2-\chi_1)\geq 0~.$$
Thus, a condition for subadditivity property for qutrits: 
\be \label{cond2} (\phi_1-\phi_2)(\chi_2-\chi_1)\geq 0.
\ee
Condition (\ref{cond2}) depends only on weights $\phi_A$ and $\phi_B$.
We derive the property (\ref{sub}) for qutrit case.

For instance, we calculate reduced matrices using formule (\ref{rm}): 
$$\psi_A \rho_A = tr_B(\phi_{AB}\rho_{AB})=tr_B\begin{pmatrix}
\phi_1 \chi_1 p_1 & 0 & 0 & 0\\
0 & \phi_1 \chi_2 p_2 & 0 & 0\\
0 & 0 & \phi_2 \chi_1 p_3 & 0\\
0 & 0 & 0 & 0
\end{pmatrix}=\begin{pmatrix}
\phi_1 \chi_1 p_1+\phi_2 \chi_1 p_3 & 0 \\
0 & \phi_1 \chi_2 p_2
\end{pmatrix},$$
$$\psi_B \rho_B = tr_A(\phi_{AB}\rho_{AB})=tr_A\begin{pmatrix}
\phi_1 \chi_1 p_1 & 0 & 0 & 0\\
0 & \phi_1 \chi_2 p_2 & 0 & 0\\
0 & 0 & \phi_2 \chi_1 p_3 & 0\\
0 & 0 & 0 & 0
\end{pmatrix}=\begin{pmatrix}
\phi_1 \chi_1 p_1+\phi_1 \chi_2 p_2 & 0 \\
0 & \phi_2 \chi_1 p_3
\end{pmatrix}.$$

Using definition of quantum weighted entropy \ref{qwe}, we can calculate  $S_{\phi_{AB}}(\rho_{AB})$, $S_{\psi_A}(\rho_A)$ and $S_{\psi_B}(\rho_B)$:
\begin{enumerate}
	\item $S_{\phi_{AB}}(\rho_{AB})=-tr\left[\begin{pmatrix}
	\phi_1 \chi_1 p_1 & 0 & 0 & 0\\
	0 & \phi_1 \chi_2 p_2 & 0 & 0\\
	0 & 0 & \phi_2 \chi_1 p_3 & 0\\
	0 & 0 & 0 & 0
	\end{pmatrix} \begin{pmatrix}
	\log p_1 & 0 & 0 & 0\\
	0 & \log p_2 & 0 & 0\\
	0 & 0 & \log p_3 & 0\\
	0 & 0 & 0 & 0
	\end{pmatrix}\right]= {} \\ {}
	=-(\phi_1 \chi_1 p_1 \log p_1+\phi_1 \chi_2 p_2 \log p_2+\phi_2 \chi_1 p_3 \log p_3);$
	\item $S_{\psi_A}(\rho_A)=-tr(\psi_A \rho_A \log \rho_A)=\begin{pmatrix}
	\phi_1 \chi_1 p_1+\phi_2 \chi_1 p_3 & 0 \\
	0 & \phi_1 \chi_2 p_2
	\end{pmatrix}\begin{pmatrix}
	\log(p_1+p_2) & 0 \\
	0 & \log p_3
	\end{pmatrix}={}\\ {}
	=-[(\phi_1 \chi_1 p_1+\phi_2 \chi_1 p_3)\log(p_1+p_2)+\phi_1 \chi_2 p_2\log p_3];$
	\item $S_{\psi_B}(\rho_B)=-[(\phi_1 \chi_1 p_1+\phi_1 \chi_2 p_2)\log(p_1+p_3)+\phi_2 \chi_1 p_3\log p_2].$
\end{enumerate}

Ineq. (2) \ref{sub} can be rewritten in form $S_{\psi_A}(\rho_A)S_{\psi_B}(\rho_B)-S_{\phi_{AB}}(\rho_{AB}) \geq 0$. After substitution from (a), (b) and (c) we obtain following expression:

$[\phi_1 \chi_1 p_1 \log p_1+\phi_1 \chi_2 p_2 \log p_2+\phi_2 \chi_1 p_3 \log p_3]-[(\phi_1 \chi_1 p_1+\phi_2 \chi_1 p_3)\log(p_1+p_2)+\phi_1 \chi_2 p_2\log p_3]-{}\\ {}
-[(\phi_1 \chi_1 p_1+\phi_1 \chi_2 p_2)\log(p_1+p_3)+\phi_2 \chi_1 p_3\log p_2]\geq 0.$

Transforming previous expression we get following inequality: 
 \be \label{ineq} -[\phi_1 \chi_1 p_1 \log [\frac{(p_1+p_2)(p_1+p_3)}{p_1}]+\phi_1 \chi_2 p_2 \log [p_1+p_2]+\phi_2 \chi_1 p_3 \log[p_1+p_3]]\geq 0. \ee

Let us consider some particular case for qutrit. We choose probabilities $p_1, p_2, p_3$ weigths $\phi_A$ and $\phi_B$ to satisfy the condition (\ref{cond2}): $$\phi_A=\begin{pmatrix}
3/4 & 0\\
0 & 1/4
\end{pmatrix},~~~~~~~~~~ \phi_B=\begin{pmatrix}
1/3 & 0\\
0 & 2/3
\end{pmatrix};$$ $$p_1=p_2=1/10,~~~~~~p_3=8/10.$$

These values satisfy the required condition (\ref{cond2}):
$$(3/4-1/4)(2/3-1/3)=1/6\geq 0.$$

Subadditivity property for this case is correct: \\
$$-(\frac{3}{4} \cdot\frac{1}{3} \cdot\frac{1}{10} \log(18/10)+\frac{3}{4} \cdot\frac{2}{3} \cdot\frac{1}{10} \log(2/10)+\frac{1}{12} \cdot\frac{8}{10} \log(9/10))=0.0728\geq 0.$$

Difference $S_{\psi_A}(\rho_A)S_{\psi_B}(\rho_B)-S_{\phi_{AB}}(\rho_{AB})$ is equal to mutual information I.

Thus, we have following expression:
{\small $$I=I(p_1, p_2, \phi_1, \phi_2, \chi_1, \chi_2)=-\left(\phi_1 \chi_1 p_1 \log \left[\frac{(p_1+p_2)(p_1+p_3)}{p_1}\right]+\phi_1 \chi_2 p_2 \log [p_1+p_2]+\phi_2 \chi_1 p_3 \log[p_1+p_3]\right)\geq0.$$}

Varying two out of six variables of mutual information $I(p_1, p_2, \phi_1, \phi_2, \chi_1, \chi_2)$ we can plot 3D graph showing dependence of mutual information on probabilities and weights. Let is consider some particular cases.

1) Weights $\phi_A=	\begin{pmatrix} 3/4 & 0 \\ 0 & 1/4 \end{pmatrix}$ and $\phi_B=	\begin{pmatrix} 
1/3 & 0 \\ 0 & 2/3 \end{pmatrix}$ satisfy the condition (\ref{cond2}) of subadditivity:
$ \begin{cases} \phi_1 - \phi_2 =3/4 - 1/4 =1/2\geq 0
\\ \chi_2-\chi_1 =2/3-1/3=1/3\geq 0\end{cases}.$ Varying probabilities $p_1$ and $p_2$ we plot dependence $I(p_1,p_2)$ illustrated on Fig.1.

2) Considering $\chi_2=1-\chi_1$, $\phi_2=1-\phi_1$, we fix probabilities $p_1, p_2, p_3$ to plot the dependence of mutual information on weights: $p_1=1/4$, $p_2=1/8$, $p_3=1-p_1-p_2$. 
This dependence $I(\phi_1,\chi_1)$ is illustrated on Fig.2 and Fig.3. Plot consists of 2 parts, according to the condition \ref{cond2},  part \textbf{a}: 
$ \begin{cases} \phi_1 - \phi_2 \geq 0\\ \chi_2-\chi_1\geq 0 \end{cases}$  and part \textbf{b} of the plot: $ \begin{cases} \phi_1 - \phi_2 \leq 0\\ \chi_2-\chi_1\leq 0 \end{cases}$ .
\begin{figure}[h!]
	\begin{center}
				\includegraphics[width=70mm]{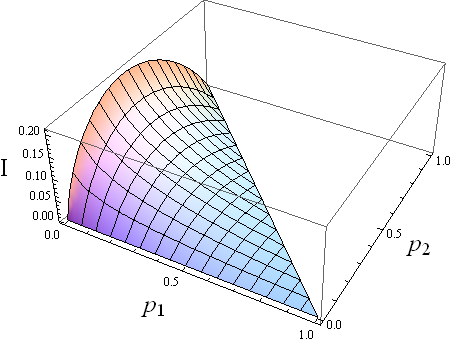}
		\hfill
		\caption{The dependence of mutual information $ I(p_1, p_2) $ on probabilities $p_1$ and $p_2$}
	\end{center}
\end{figure}
\begin{figure}[h!]
	\begin{multicols}{2}
		\hfill
		\includegraphics[width=70mm]{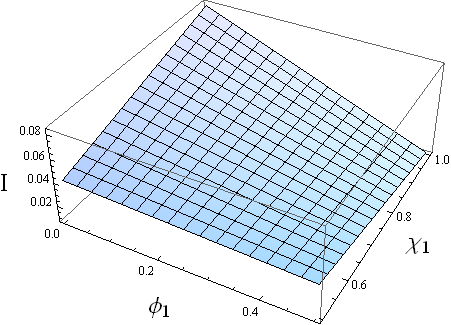}
		\hfill
		\caption{The dependence of mutual information $ I(\phi_1, \chi_1) $ on weights $\phi_1$ and $\chi_1$: third case \textbf{a}}
		\label{figLeft}
		\hfill
		\includegraphics[width=70mm]{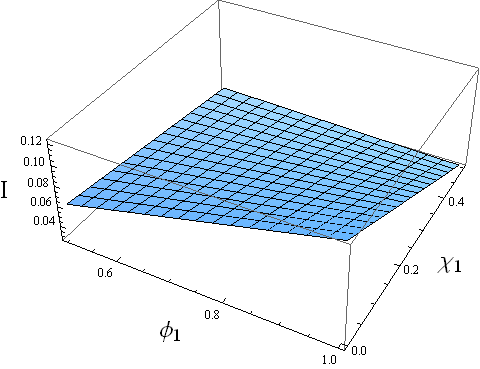}
		\hfill
		\caption{The dependence of mutual information $ I(\phi_1, \chi_1) $ on weights $\phi_1$ and $\chi_1$: third case \textbf{b}}
		\label{figRight}
	\end{multicols}
\end{figure}

\newpage\section{Conclusion}

To resume, we point out two main results of our work.
Using the approach developed in \cite{suhov} we extended the notion of weighted entropies known for multipartite system states and entropies of its subsystem states to the case of single qudit. We showed that the subadditivity condition for the weighted entropies of bipartite system obtained in \cite{suhov} can be formulated also for the weighted entropies of the noncomposite system like qutrit state. The new inequality can be  checked experimentally for superconductive circuit states discussed, e.g. in \cite{kikt1, kikt2, glush}. Other new entropic inequalities for weighted entropy, e.g. strong subadditivity condition found in \cite{suhov} for tripartite systems, can be considered and extended to the case of noncomposite system. Applying the nonlinear quantum channel to density matrix $\rho_{AB}$ defined in (\ref{matrix}), we get the transformed density $4 \times 4$ matrix of the form $\hat{\rho}'=\frac{\hat{P}\hat{\rho}\hat{P}}{Tr \hat{P}\hat{\rho}\hat{P}}$ (defined by Eq. (10) in \cite{europhys}). 
 For chosen rank-2 projector 
 $\hat{P}=\begin{pmatrix}
 1 & 0 & 0 & 0\\
 0 & 0 & 0 & 0\\
 0 & 0 & 1 & 0\\
 0 & 0 & 0 & 0
 \end{pmatrix}$, and therefore 
 $\hat{\rho_{AB}}'=\begin{pmatrix}
 \frac{p_1}{p_1+p_3} & 0 & 0 & 0\\
 0 & 0 & 0 & 0\\
 0 & 0 & \frac{p_3}{p_1+p_3} & 0\\
 0 & 0 & 0 & 0
 \end{pmatrix}$,
we can get transformed weighted entropy inequality (\ref{ineq}) which reads 
 $\phi_1 p_1 + \phi_2 p_2 \geq 0  $.
 \\ This obvious example can be extended to other states of qudits to get more complicated inequalities. We study such examples in future publications.
 
\section{Acknowledgments}
Formulation of the problem and results of sections I and II are obtained by V. I. Man'ko, supported by Russian Science Foundation under Project No. 16-11-00084 and performed in Moscow Institute of Physics and Technology.

\end{document}